\documentclass[12pt,centertags,dvips,twoside]{report}
\makeatletter
\def\include#1{\relax
  \ifnum\@auxout=\@partaux
    \@latex@error{\string\include\space cannot be nested}\@eha
  \else \@include#1 \fi}
\def\@include#1 {%
  \clearpage
  \if@filesw
    \immediate\write\@mainaux{\string\@input{#1.aux}}%
  \fi
  \@tempswatrue
  \if@partsw
    \@tempswafalse
    \edef\reserved@b{#1}%
    \@for\reserved@a:=\@partlist\do
      {\ifx\reserved@a\reserved@b\@tempswatrue\fi}%
  \fi
  \if@tempswa
    \let\@auxout\@partaux
    \if@filesw
      \immediate\openout\@partaux #1.aux
      \immediate\write\@partaux{\relax}%
    \fi
    \@input@{#1.inc}%
    \clearpage
    \@writeckpt{#1}%
    \if@filesw
      \immediate\closeout\@partaux
    \fi
  \else
    \deadcycles\z@
    \@nameuse{cp@#1}%
  \fi
  \let\@auxout\@mainaux}
\makeatother
\usepackage{dsfont}
\usepackage{mathrsfs}
\usepackage{latexsym}
\usepackage{mcite}
\usepackage{konstruktion}
\PapierFormat <210mm,297mm>
\Aehnlichkeitskonstruktion <16cm>
\usepackage{fancyhdr}
\fancypagestyle{plain}{%
  \fancyhf{}
  \renewcommand{\headrulewidth}{0pt}
} 
\newcommand{\clearemptydoublepage}{%
    \newpage{\pagestyle{empty}\cleardoublepage}}
\usepackage{amssymb}
\usepackage[centertags]{amsmath}
\allowdisplaybreaks[4]
\newlength{\diracchlen}
\newlength{\vecchlen}
\newlength{\vecchhgt}

\usepackage{epsfig}
\usepackage{axodraw}
\begin{document}
%
\psfull
%
\addtocontents{toc}{%
  \protect\markboth{\sf\contentsname}{\sf\contentsname}}
\addtocontents{lof}{%
  \protect\markboth{\sf\listfigurename}{\sf\listfigurename}}
\addtocontents{lot}{%
  \protect\markboth{\sf\listtablename}{\sf\listtablename}}
%
%
\pagenumbering{roman}
%
\begin{titlepage} 
\begin{flushright}
DO-TH 2000/09\\
May 2000
\end{flushright}
\begin{center}
  \vspace*{\stretch{1.5}}
  {\Huge\bf
    NLO QCD Corrections to the Polarized Photo- and
    Hadroproduction of Heavy Quarks
  \par}
  \vspace*{\stretch{3}}
  {\large
    {\bf Dissertation}\\
    zur Erlangung des Grades eines\\
    Doktors der Naturwissenschaften\\
    der Abteilung Physik\\    
    der Universit\"at Dortmund
  \par}
  \vspace*{\stretch{2}}
  {\large
    {\normalsize{} vorgelegt von}\\
    {\bf Ingo Bojak}
  \par}
  \vspace*{1em}
  {\large
    April 2000
  \par}
\end{center} \end{titlepage}
\clearemptydoublepage
%
{\thispagestyle{empty}
  \vspace*{\stretch{1}}
  \begin{center}
    {\bf\large F\"ur meine Eltern und meinen Bruder}\\
    \vspace*{5cm}
   \begin{minipage}{10.5cm}
    \begin{flushleft}
    {\em From the side, a whole range; from the end, a single peak;\\
          far, near, high, low, no two parts alike.\\
          Why can't I tell the true shape of Lu-shan?\\
          Because I myself am in the mountain.\\}
    \end{flushleft}
    \begin{flushright}
    Su Tung-p'o, 1084 A.D.,\\
    transl.\ B.\ Watson
    \end{flushright}
   \end{minipage}\\
  \end{center}  
  \vspace*{\stretch{2}}
}
\clearemptydoublepage
%
\renewcommand{\chaptermark}[1]{%
  \markboth{\chaptername\ \thechapter.\ #1}{}
  \markright{\chaptername\ \thechapter.\ #1}}
\renewcommand{\sectionmark}[1]{\markright{\thesection.\ #1}}
%
\tableofcontents
\clearpage
%
\listoffigures
\clearpage
%
\listoftables
\clearemptydoublepage
%
\pagenumbering{arabic}
%
\include{intro}
\clearpage
%
\renewcommand{\chaptermark}[1]{%
  \markboth{\chaptername\ \thechapter.\ #1}{}}
\renewcommand{\sectionmark}[1]{\markright{\thesection.\ #1}}
\include{born}
\clearpage
\include{tensor}
\clearpage
\include{renorm}
\clearpage
\include{reell}
\clearpage
%
\renewcommand{\chaptermark}[1]{%
  \markboth{\chaptername\ \thechapter.\ #1}{}
  \markright{\chaptername\ \thechapter.\ #1}}
\renewcommand{\sectionmark}[1]{\markright{\thesection.\ #1}}
\include{fakt}
\clearpage
%
\renewcommand{\chaptermark}[1]{%
  \markboth{\chaptername\ \thechapter.\ #1}{}}
\renewcommand{\sectionmark}[1]{\markright{\thesection.\ #1}}
\include{part}
\clearpage
\include{hadres}
\clearpage
%
\renewcommand{\chaptermark}[1]{%
  \markboth{\chaptername\ \thechapter.\ #1}{}
  \markright{\chaptername\ \thechapter.\ #1}}
\renewcommand{\sectionmark}[1]{\markright{\thesection.\ #1}}
\include{sum}
\clearemptydoublepage
\appendix
\renewcommand{\chaptermark}[1]{%
  \markboth{Appendix\ \thechapter.\ #1}{}}
\renewcommand{\sectionmark}[1]{\markright{\thesection.\ #1}}
\include{app_virt}
\clearpage
\include{tricks}
\clearpage
\renewcommand{\chaptermark}[1]{%
  \markboth{Appendix\ \thechapter.\ #1}{}
  \markright{Appendix\ \thechapter.\ #1}}
\include{viso}
\clearemptydoublepage
%
\bibliographystyle{mydoc}
\fancyhead[RE]{\sf\nouppercase{\leftmark}}
\fancyhead[LO]{\sf\nouppercase{\rightmark}}
\bibliography{hadro}

\begin{thebibliography}{100}
\expandafter\ifx\csname url\endcsname\relax
  \def\url#1{\texttt{#1}}\fi
\expandafter\ifx\csname urlprefix\endcsname\relax\def\urlprefix{URL }\fi

\bibitem{cit:pdg}
C.~{Caso et al., Particle Data Group}, Eur. Phys. J. \textbf{C3} (1998) 1.

\bibitem{cit:atom}
V.~S. Shirley, C.~M. Baglin, S.~Y.~F. Chu and J.~Zipkin, eds., \emph{Table of
  Isotopes, 8th Ed.}, John Wiley \& Sons, 1996.

\bibitem{cit:revlead}
M.~Anselmino, A.~Efremov and E.~Leader, Phys. Rep. \textbf{261} (1995) 1.
  Erratum: {\bf 281} (1997) 399.

\bibitem{cit:revcheng}
H.-Y. Cheng, Int. J. Mod. Phys. \textbf{A11} (1996) 5109.

\bibitem{cit:revvett}
M.~C. Vetterli, in \emph{Quantum Chromodynamics}, edited by A.~Astbury, B.~A.
  Campbell, F.~C. Khanna, J.~L. Pinfold and M.~Vetterli, p.~49, Proceedings of
  the {\em Lake Louise Winter Institute}, Lake Louise, Canada, World
  Scientific, 1998.

\bibitem{cit:revreya}
B.~Lampe and E.~Reya, \emph{Spin Physics and Polarized Structure Functions},
  Report DO-TH 98/02 and MPI-PhT/98-23, Universit{\"a}t Dortmund and
  Max-Planck-Institut f{\"u}r Physik, 1998. {\tt hep-ph/9810270}.

\bibitem{cit:close}
F.~E. Close, \emph{An Introduction to Quarks and Partons}, Academic Press,
  1979.

\bibitem{cit:roclo}
F.~E. Close and R.~G. Roberts, Phys. Lett. \textbf{B316} (1993) 165.

\bibitem{cit:sehgal}
L.~M. Sehgal, Phys. Rev. \textbf{D10} (1974) 1663. Erratum: {\bf D11} (1975)
  2016.

\bibitem{cit:mano}
R.~L. Jaffe and A.~Manohar, Nucl. Phys. \textbf{B337} (1990) 509.

\bibitem{cit:gourdin}
M.~Gourdin, Nucl. Phys. \textbf{B38} (1972) 418.

\bibitem{cit:ej}
J.~Ellis and R.~L. Jaffe, Phys. Rev. \textbf{D9} (1974) 1444. Erratum: {\bf
  D10} (1974) 1669.

\bibitem{cit:bj1}
J.~D. Bj{\o}rken, Phys. Rev. \textbf{148} (1966) 1467.

\bibitem{cit:bj2}
J.~D. Bj{\o}rken, Phys. Rev. \textbf{D1} (1970) 1376.

\bibitem{cit:emc1}
J.~{Ashman et al., EM Collab.}, Phys. Lett. \textbf{B206} (1988) 364.

\bibitem{cit:emc2}
J.~{Ashman et al., EM Collab.}, Nucl. Phys. \textbf{B328} (1989) 1.

\bibitem{cit:e143a}
K.~{Abe et al., E143 Collab.}, Phys. Rev. Lett. \textbf{74} (1995) 346.

\bibitem{cit:e143b}
K.~{Abe et al., E143 Collab.}, Phys. Rev. Lett. \textbf{75} (1995) 25.

\bibitem{cit:bjsmc}
B.~{Adeva et al., SM Collab.}, Phys. Lett. \textbf{B412} (1997) 414.

\bibitem{cit:hbj1}
J.~Kodaira, S.~Matsuda, K.~Sasaki and T.~Uematsu, Nucl. Phys. \textbf{B159}
  (1979) 99.

\bibitem{cit:hbj2}
J.~Kodaira, S.~Matsuda, M.~Muta, K.~Sasaki and T.~Uematsu, Phys. Rev.
  \textbf{D20} (1979) 627.

\bibitem{cit:hbj3}
S.~A. Larin and J.~A.~M. Vermaseren, Phys. Lett. \textbf{B259} (1991) 345.

\bibitem{cit:grsv}
M.~Gl{\"u}ck, E.~Reya, M.~Stratmann and W.~Vogelsang, Phys. Rev. \textbf{D53}
  (1996) 4775.

\bibitem{cit:newf1}
E.~Leader, A.~V. Sidorov and D.~B. Stamenov, Phys. Lett. \textbf{B462} (1999)
  189.

\bibitem{cit:newf2}
Y.~{Goto et al., AA Collab.}, \emph{Polarized Parton Distribution Functions
  from the $A_1$ Asymmetry Data}, Report RIKEN-AF-NP-324, SAGA-HE-143-99,
  KOBE-FHD-99-03 and FUT-99-02, The Institute of Physical and Chemical Research
  (RIKEN) and other institutions, 1999. {\tt hep-ph/0001046}.

\bibitem{cit:dss}
D.~de~Florian, O.~A. Sampayo and R.~Sassot, Phys. Rev. \textbf{D57} (1998)
  5803.

\bibitem{cit:gs}
T.~Gehrmann and W.~J. Stirling, Phys. Rev. \textbf{D53} (1996) 6100.

\bibitem{cit:smc}
B.~{Adeva et al., SM Collab.}, Phys. Rev. \textbf{D58} (1998) 112002.

\bibitem{cit:abs}
R.~D. Ball, S.~Forte and G.~Ridolfi, Phys. Lett. \textbf{B378} (1996) 255.

\bibitem{cit:klein}
M.~Klein, \emph{Structure Functions in Deep Inelastic Lepton-Nucleon
  Scattering}, Talk given at the {\em ``19th International Symposium on Lepton
  and Photon Interactions at High-Energies, LP 99''}, Stanford, USA, 9-14 Aug
  1999. {\tt hep-ex/0001059}.

\bibitem{cit:h70}
C.~{Adloff et al., H1 Collab.}, Nucl. Phys. \textbf{B545} (1999) 21.

\bibitem{cit:vv}
W.~Vogelsang and A.~Vogt, Nucl. Phys. \textbf{B453} (1995) 334.

\bibitem{cit:cteq1}
J.~{Huston et al.}, Phys. Rev. \textbf{D58} (1998) 114034.

\bibitem{cit:cteq2}
H.~L. {Lai et al., CTEQ Collab.}, Eur. Phys. J. \textbf{C12} (2000) 375.

\bibitem{cit:mrst1}
A.~D. Martin, R.~G. Roberts, W.~J. Stirling and R.~S. Thorne, Eur. Phys. J.
  \textbf{C4} (1998) 463.

\bibitem{cit:grv98}
M.~Gl{\"u}ck, E.~Reya and A.~Vogt, Eur. Phys. J. \textbf{C5} (1998) 461.

\bibitem{cit:mrst2}
A.~D. Martin, R.~G. Roberts, W.~J. Stirling and R.~S. Thorne, \emph{Parton
  Distributions and the LHC: $W$ and $Z$ Production}, Report DTP-99-64, Durham
  University, 1999.

\bibitem{cit:spexp1}
J.~Bl{\"u}mlein, A.~{De Roeck}, T.~Gehrmann and W.-D. Nowak, eds., \emph{{\em
  Proceedings of the Workshop} ``Deep Inelastic Scattering off Polarized
  Targets'', {\em Zeuthen, Germany}}, DESY, 1997. Report DESY 97-200.

\bibitem{cit:spexp2}
E.~W. Hughes and R.~Voss, Annu. Rev. Nucl. Part. Sci. \textbf{49} (1999) 303.

\bibitem{cit:nps1}
R.~Mertig and W.~L. van Neerven, Z. Phys. \textbf{C70} (1996) 637.

\bibitem{cit:nlospl1}
W.~Vogelsang, Phys. Rev. \textbf{D54} (1996) 2023.

\bibitem{cit:nlospl2}
W.~Vogelsang, Nucl. Phys. \textbf{B475} (1996) 47.

\bibitem{cit:herapol}
A.~{De Roeck} and T.~Gehrmann, eds., \emph{{\em Proceedings of the Workshop}
  ``Physics with Polarized Protons at HERA'', {\em Hamburg and Zeuthen,
  Germany}}, DESY, 1997. Report DESY 97-233.

\bibitem{cit:hermes}
A.~{Airapetian et al., HERMES Collab.}, \emph{Measurement of the Spin Asymmetry
  in the Photoproduction of Pairs of High-$p_T$ Hadrons at HERMES}, Report
  DESY-99-071, DESY, 1999. To appear in Phys. Rev. Lett.

\bibitem{cit:dead}
D.~de~Florian, M.~Stratmann and W.~Vogelsang, \emph{private communication}.

\bibitem{cit:com1}
G.~{Baum et al., COMPASS Collab.}, \emph{Common Muon and Proton Apparatus for
  Structure and Spectroscopy}, Report CERN/SPSLC 96-14, CERN, 1996.

\bibitem{cit:com2}
G.~{Baum et al., COMPASS Collab.}, \emph{Common Muon and Proton Apparatus for
  Structure and Spectroscopy -- Addendum 1}, Report CERN/SPSLC 96-30, CERN,
  1996.

\bibitem{cit:rhic1}
D.~{Hill et al., RHIC Spin Collab.}, \emph{Letter of Intent}, Report
  RHIC-SPIN-LOI-1991, updated 1993, RIKEN BNL Research Center, 1991.

\bibitem{cit:rhic2}
G.~{Bunce et al., RHIC Spin Collab.}, Particle World \textbf{3} (1992) 1.

\bibitem{cit:rhic3}
\emph{{\em Proceedings of the} RSC Annual Meeting, {\em Marseille, France}},
  Centre de Physique Th{\'e}orique, 1996. CPT-96/P.3400.

\bibitem{cit:rhic4}
\emph{{\em Proceedings of the 1998/2000 Workshops} ``RHIC Spin Physics'' {\em
  /} ``Predictions and Uncertainties for RHIC Spin Physics'', {\em RIKEN BNL
  Research Center, New York, USA}}, Brookhaven National Laboratory, to appear.

\bibitem{cit:svh1}
M.~Stratmann and W.~Vogelsang, in \emph{Future Physics at HERA}, edited by
  G.~Ingelman, A.~{De Roeck} and R.~Klanner, p. 815, Proceedings of the 1995/96
  Workshop {\em ``Future Physics at HERA''}, Hamburg, Germany, DESY, 1996.

\bibitem{cit:svh2}
M.~Stratmann and W.~Vogelsang, Z. Phys. \textbf{C74} (1997) 641.

\bibitem{cit:letter}
I.~Bojak and M.~Stratmann, Phys. Lett. \textbf{B433} (1998) 411.

\bibitem{cit:long}
I.~Bojak and M.~Stratmann, Nucl. Phys. \textbf{B540} (1999) 345.

\bibitem{cit:msbp}
L.~E. Gordon and W.~Vogelsang, Phys. Rev. \textbf{D48} (1993) 3136.

\bibitem{cit:erratum}
I.~Bojak and M.~Stratmann, Erratum for Nucl. Phys. {\bf B540} (1999) 345: {\bf
  B569} (2000) 694 and Erratum for Phys. Lett. {\bf B433} (1998) 411: to be
  published.

\bibitem{cit:dis99}
I.~Bojak, in \emph{DIS '99}, edited by J.~Bl{\"u}mlein and T.~Riemann, p. 599,
  Proceedings of the {\em ``7th International Workshop on Deep Inelastic
  Scattering and QCD, DIS '99''}, Zeuthen, Germany, 1999,. Nucl. Phys. {\bf B}
  (Proc. Suppl.) {\bf 79} (1999) October 1999.

\bibitem{cit:gr}
M.~Gl{\"u}ck and E.~Reya, Z. Phys. \textbf{C39} (1988) 569.

\bibitem{cit:oth1}
G.~Altarelli and W.~J. Stirling, Particle World \textbf{1} (1989) 40.

\bibitem{cit:oth2}
M.~Gl{\"u}ck, E.~Reya,  and W.~Vogelsang, Nucl. Phys. \textbf{B351} (1991) 579.

\bibitem{cit:oth3}
S.~I. Alekhin, V.~I. Borodulin and S.~F. Sultanov, Int. J. Mod. Phys.
  \textbf{A8} (1993) 1603.

\bibitem{cit:oth4}
S.~Keller and J.~F. Owens, Phys. Rev. \textbf{D49} (1994) 1199.

\bibitem{cit:oth5}
S.~Frixione and G.~Ridolfi, Phys. Lett. \textbf{B383} (1996) 227.

\bibitem{cit:lopol}
A.~P. Contogouris, S.~Papadopoulos and B.~Kamal, Phys. Lett. \textbf{B246}
  (1990) 523.

\bibitem{cit:lopo2}
M.~Karliner and R.~W. Robinett, Phys. Lett. \textbf{B324} (1994) 209.

\bibitem{cit:naspho}
R.~K. Ellis and P.~Nason, Nucl. Phys. \textbf{B312} (1988) 551.

\bibitem{cit:nashad}
S.~Dawson, R.~K. Ellis and P.~Nason, Nucl. Phys. \textbf{B303} (1988) 607.

\bibitem{cit:bkns}
W.~J.~P. Beenakker, H.~Kuijf, W.~L. van Neerven and J.~Smith, Phys. Rev.
  \textbf{D40} (1989) 54.

\bibitem{cit:smith3}
W.~J.~P. Beenakker, W.~L. van Neerven, R.~Meng, G.~A. Schuler and J.~Smith,
  Nucl. Phys. \textbf{B351} (1991) 507.

\bibitem{cit:smne}
J.~Smith and W.~L. van Neerven, Nucl. Phys. \textbf{B374} (1992) 36.

\bibitem{cit:passvelt}
G.~Passarino and W.~Veltman, Nucl. Phys. \textbf{B160} (1979) 151.

\bibitem{cit:cragie}
N.~S. Craigie, K.~Hidaka, M.~Jacob and F.~M. Renard, Phys. Rep. \textbf{99},
  Nos. 2 \& 3 (1983) 69.

\bibitem{cit:ghost}
J.~Babcock, D.~Sivers and S.~Wolfram, Phys. Rev. \textbf{D18} (1978) 162.

\bibitem{cit:chli}
T.-P. Cheng and L.-F. Li, \emph{Gauge Theory of Elementary Particle Physics},
  Clarendon Press, 1984.

\bibitem{cit:sterman}
G.~Sterman, \emph{An Introduction to Quantum Field Theory}, Cambridge
  University Press, 1993.

\bibitem{cit:slav1}
G.~'t~Hooft, Nucl. Phys. \textbf{B33} (1971) 173.

\bibitem{cit:slav2}
J.~C. Taylor, Nucl. Phys. \textbf{B33} (1971) 436.

\bibitem{cit:hvbm1}
G.~'t~Hooft and M.~Veltman, Nucl. Phys. \textbf{B44} (1972) 189.

\bibitem{cit:hvbm2}
P.~Breitenlohner and D.~Maison, Comm. Math. Phys. \textbf{52} (1977) 11.

\bibitem{cit:gott}
K.~Gottfried and J.~D. Jackson, Nuovo Cim. \textbf{34} (1964) 735.

\bibitem{cit:ellis}
R.~K. Ellis, M.~A. Furman, H.~E. Haber and I.~Hinchliffe, Nucl. Phys.
  \textbf{B173} (1980) 397.

\bibitem{cit:math}
S.~Wolfram, \emph{{\tt Mathematica} -- Ver. 3 or higher}, Wolfram Research,
  1997.

\bibitem{cit:tracer}
M.~Jamin and M.~E. Lautenbacher, Comp. Phys. Comm. \textbf{74} (1993) 265.

\bibitem{cit:lounp}
L.~M. Jones and H.~W. Wyld, Phys. Rev. \textbf{D17} (1978) 759.

\bibitem{cit:ngg1}
B.~Kamal, Z.~Merebashvili and A.~P. Contogouris, Phys. Rev. \textbf{D51} (1995)
  4808.

\bibitem{cit:ngg2}
G.~Jikia and A.~Tkabladze, Phys. Rev. \textbf{D54} (1996) 2030.

\bibitem{cit:gor1}
M.~Gl{\"u}ck, J.~F. Owens and E.~Reya, Phys. Rev. \textbf{D17} (1978) 2324.

\bibitem{cit:gor2}
B.~L. Combridge, Nucl. Phys. \textbf{B151} (1979) 429.

\bibitem{cit:grv94}
M.~Gl{\"u}ck, E.~Reya and A.~Vogt, Z. Phys. \textbf{C67} (1995) 433.

\bibitem{cit:mallot}
G.~K. Mallot, \emph{Accessing the Gluon Polarisation in Deep Inelastic Muon
  Scattering}, {\tt http://www.compass.cern.ch/compass/sp/ps/erice\_95.ps.gz}.
  Talk given at the Conference {\em ``Quarks in Hadrons and Nuclei''}, Erice,
  Italy, 19-27 Sept. 1995.

\bibitem{cit:muta}
T.~Muta, \emph{Foundations of Quantum Chromodynamics}, World Scientific, 1987.

\bibitem{cit:nowak}
M.~A. Nowak, M.~Praszalowicz and W.~Slomi\'{n}ski, Ann. Phys. (NY) \textbf{166}
  (1986) 443.

\bibitem{cit:abramowitz}
M.~Abramowitz and I.~A. Stegun, eds., \emph{Handbook of Mathematical
  Functions}, National Bureau of Standards, 1964.

\bibitem{cit:docbeen}
W.~J.~P. Beenakker, \emph{Electroweak Corrections: Techniques and
  Applications}, Ph.D. thesis, University of Leiden, 1989.

\bibitem{cit:pokorski}
S.~Pokorski, \emph{Gauge Field Theories}, Cambridge University Press, 1987.

\bibitem{cit:tarrach}
P.~Pascual and R.~Tarrach, \emph{QCD: Renormalization for the Practitioner},
  Springer, 1984.

\bibitem{cit:rvsc1}
F.~Bloch and A.~Nordsieck, Phys. Rev. \textbf{52} (1937) 54.

\bibitem{cit:rvsc2}
A.~Nordsieck, Phys. Rev. \textbf{52} (1937) 59.

\bibitem{cit:rvsc3}
T.~Kinoshita, J. Math. Phys. \textbf{3} (1962) 650.

\bibitem{cit:rvsc4}
T.~D. Lee and M.~Nauenberg, Phys. Rev. \textbf{133} (1964) B1549.

\bibitem{cit:qian}
S.~Qian, \emph{A New Renormalization Prescription (CWZ Subtraction Scheme) for
  QCD and its Application to DIS}, Report ANL-HEP-PR-84-72, Argonne National
  Laboratory, 1984.

\bibitem{cit:slav3}
A.~A. Slavnov, Teor. Mat. Fiz. \textbf{10} (1972) 153. [Theor.\ Math.\ Phys.\
  {\bf 10} (1973) 99].

\bibitem{cit:ward1}
J.~C. Ward, Phys. Rev. \textbf{78} (1950) 182.

\bibitem{cit:ward2}
Y.~Takahashi, Nuovo Cim. \textbf{6} (1957) 371.

\bibitem{cit:collins}
J.~C. Collins, \emph{Renormalization}, Cambridge University Press, 1984.

\bibitem{cit:brs}
C.~Becchi, A.~Rouet and R.~Stora, Ann. Phys. (NY) \textbf{98} (1976) 287.

\bibitem{cit:cwz}
J.~C. Collins, F.~Wilczek and A.~Zee, Phys. Rev. \textbf{D18} (1978) 242.

\bibitem{cit:colfak}
J.~C. Collins, Phys. Rev. \textbf{D58} (1998) 094002.

\bibitem{cit:ditr}
D.~Coleman and D.~Gross, Phys. Rev. Lett. \textbf{31} (1973) 851.

\bibitem{cit:be1}
D.~R.~T. Jones, Nucl. Phys. \textbf{B75} (1974) 531.

\bibitem{cit:be2}
W.~E. Caswell, Phys. Rev. Lett. \textbf{33} (1974) 244.

\bibitem{cit:be3}
{\'{E}.~{Sh.}~Egoryan} and O.~V. Tarasov, Teor. Mat. Fiz. \textbf{41} (1979)
  26. [Theor.\ Math.\ Phys.\ {\bf 41} (1979) 863].

\bibitem{cit:tarma}
R.~Tarrach, Nucl. Phys. \textbf{B183} (1981) 384.

\bibitem{cit:been}
W.~J.~P. Beenakker, \emph{private communication}.

\bibitem{cit:gradstein}
I.~S. Gradshteyn and I.~M. Ryzhik, \emph{Table of Integrals, Series, and
  Products --- Corrected and Enlarged Edition}, Academic Press, 1980.

\bibitem{cit:neecoll}
W.~L. van Neerven, Nucl. Phys. \textbf{B268} (1986) 453.

\bibitem{cit:mfak}
R.~K. Ellis, H.~Georgi, M.~Machacek, H.~D. Politzer and G.~G. Ross, Nucl. Phys.
  \textbf{B152} (1979) 285.

\bibitem{cit:altp1}
G.~Altarelli and G.~Parisi, Nucl. Phys. \textbf{B126} (1977) 298.

\bibitem{cit:altp2}
M.~A. Ahmed and G.~G. Ross, Nucl. Phys. \textbf{B111} (1976) 441.

\bibitem{cit:altp3}
V.~N. Gribov and L.~N. Lipatov, Yad. Fiz. \textbf{15} (1972) 781 and 1218.
  [Sov. J. Nucl. Phys. {\bf 15} (1972) 438 and 675].

\bibitem{cit:altp4}
{Yu.~L.~Dokshitzer}, Zh. Eksp. Teor. Fiz \textbf{73} (1977) 1216. [Sov. Phys.
  JETP {\bf 46} (1977) 641].

\bibitem{cit:halzen}
F.~Halzen and A.~D. Martin, \emph{Quarks and Leptons}, John Wiley \& Sons,
  1984.

\bibitem{cit:paige}
F.~E. Paige, \emph{QCD and Event Simulation}, Report BNL-43525, Brookhaven
  National Laboratory, Lectures given at the {\em Theoretical Advanced Summer
  Institute}, Boulder, USA, 1989.

\bibitem{cit:dipl}
I.~Bojak, \emph{Gluonen und Strukturfunktionen bei kleinen Bj{\o}rken $x$}.
  Diploma Thesis, Universit{\"a}t Dortmund, Germany, 1995 (unpublished).

\bibitem{cit:nus1}
E.~G. Floratos, D.~A. Ross and C.~T. Sachrajda, Nucl. Phys. \textbf{B129}
  (1977) 66. Erratum: {\bf B139} (1978) 545.

\bibitem{cit:nus2}
E.~G. Floratos, D.~A. Ross and C.~T. Sachrajda, Nucl. Phys. \textbf{B152}
  (1979) 493.

\bibitem{cit:nus3}
A.~Gonz\'{a}lez-Arroyo, C.~L\'{o}pez and F.~J. Yndur\'{a}in, Nucl. Phys.
  \textbf{B153} (1979) 161.

\bibitem{cit:nus4}
A.~Gonz\'{a}lez-Arroyo and C.~L\'{o}pez, Nucl. Phys. \textbf{B166} (1980) 429.

\bibitem{cit:nus6}
G.~Curci, W.~Furmanski and R.~Petronzio, Nucl. Phys. \textbf{B175} (1980) 27.

\bibitem{cit:nus7}
W.~Furmanski and R.~Petronzio, Phys. Lett. \textbf{B97} (1980) 437.

\bibitem{cit:nus8}
E.~G. Floratos, C.~Kounnas and R.~Lacaze, Nucl. Phys. \textbf{B192} (1981) 417.

\bibitem{cit:nus11}
W.~Furmanski and R.~Petronzio, Z. Phys. \textbf{C11} (1982) 293.

\bibitem{cit:nus10}
R.~Hamberg and W.~L. van Neerven, Nucl. Phys. \textbf{B379} (1992) 143.

\bibitem{cit:ober}
F.~Oberhettinger, \emph{Tables of Mellin Transforms}, Springer, 1974.

\bibitem{cit:mod1}
M.~Gl{\"u}ck and W.~Vogelsang, Z. Phys. \textbf{C55} (1992) 353.

\bibitem{cit:mod2}
M.~Gl{\"u}ck and W.~Vogelsang, Z. Phys. \textbf{C57} (1993) 309.

\bibitem{cit:mod3}
M.~Gl{\"u}ck, M.~Stratmann and W.~Vogelsang, Phys. Lett. \textbf{B337} (1994)
  373.

\bibitem{cit:grvphot}
M.~Gl{\"u}ck, E.~Reya and A.~Vogt, Phys. Rev. \textbf{D46} (1992) 1973.

\bibitem{cit:mod4}
M.~Gl{\"u}ck, E.~Reya and I.~Schienbein, Phys. Rev. \textbf{D60} (1999) 054019.

\bibitem{cit:svphot}
M.~Stratmann and W.~Vogelsang, Phys. Lett. \textbf{B386} (1996) 370.

\bibitem{cit:grvdis}
M.~Gl{\"u}ck, E.~Reya and A.~Vogt, Phys. Rev. \textbf{D45} (1992) 3986.

\bibitem{cit:delta1}
J.~Kubar-Andr{\'e} and F.~E. Paige, Phys. Rev. \textbf{D19} (1979) 221.

\bibitem{cit:delta2}
B.~Humpert and W.~L. van Neerven, Nucl. Phys. \textbf{B184} (1981) 225.

\bibitem{cit:neben}
D.~de~Florian, S.~Frixione, A.~Signer and W.~Vogelsang, Nucl. Phys.
  \textbf{B539} (1999) 455.

\bibitem{cit:yorgos}
G.~Tsipolitis, \emph{private communication}.

\bibitem{cit:nrqcd1}
E.~Braaten, B.~A. Kniehl and J.~Lee, \emph{Polarization of prompt $J/\psi$ at
  the Tevatron}, Report DESY-99-175, DESY, 1999. {\tt hep-ph/9911436}.

\bibitem{cit:nrqcd2}
G.~T. Bodwin, E.~Braaten and G.~P. Lepage, Phys. Rev. \textbf{D51} (1995) 1125.

\bibitem{cit:cem1}
J.~F. Amundson, O.~J.~P. {\'E}boli, E.~M. Gregores and F.~Halzen, Phys. Lett.
  \textbf{B372} (1996) 127.

\bibitem{cit:cem2}
J.~F. Amundson, O.~J.~P. {\'E}boli, E.~M. Gregores and F.~Halzen, Phys. Lett.
  \textbf{B390} (1997) 323.

\bibitem{cit:cdfjpsi}
A.~{Sansoni et al., CDF Collab.}, Nuovo Cim. \textbf{109A} (1996) 827.

\bibitem{cit:ww1}
C.~F. von Weizs{\"a}cker, Z. Phys. \textbf{88} (1934) 612.

\bibitem{cit:ww2}
E.~J. Williams, Phys. Rev. \textbf{45} (1934) 729.

\bibitem{cit:ww3}
S.~Frixione, M.~L. Mangano, P.~Nason and G.~Ridolfi, Phys. Lett. \textbf{B319}
  (1993) 339.

\bibitem{cit:ww4}
D.~de~Florian and S.~Frixione, Phys. Lett. \textbf{B457} (1999) 236.

\bibitem{cit:pet}
C.~Peterson, D.~Schlatter, I.~Schmitt and P.~M. Zerwas, Phys. Rev. \textbf{D27}
  (1983) 105.

\bibitem{cit:e691}
J.~C. {Anjos et al., E691 Collab.}, Phys. Rev. Lett. \textbf{62} (1989) 513.

\bibitem{cit:na14-2}
M.~P. {Alvarez et al., NA14/2 Collab.}, Z. Phys. \textbf{C60} (1993) 53.

\bibitem{cit:e687}
G.~Bellini, in \emph{Results and Perspective in Particle Physics}, edited by
  M.~Greco, p. 435, Proceedings of {\em ``Les Rencontres de Physique de la
  Vall{\'e} d'Aoste''}, La Thuille, France, Editions Fronti{\`e}res, 1994.

\bibitem{cit:sam1}
S.~Frixione, M.~L. Mangano, P.~Nason and G.~Ridolfi, in \emph{Heavy Flavours
  II}, edited by A.~J. Buras and M.~Lindner, Advanced Series on Directions in
  High Energy Physics -- Vol. 15, p. 609, World Scientific, 1998.

\bibitem{cit:beam1}
E.~Norrbin and T.~Sj{\"o}strand, Phys. Lett. \textbf{B442} (1998) 407.

\bibitem{cit:brav}
A.~Bravar, D.~von Harrach and A.~Kotzinian, Phys. Lett. \textbf{B421} (1998)
  349.

\bibitem{cit:harris}
B.~W. Harris, \emph{Open Charm Production in Deep Inelastic Scattering at
  Next-to-Leading Order at HERA}, Report ANL-HEP-CP-99-69, Argonne National
  Laboratory, 1999. To appear in the Proceedings of the Ringberg Workshop {\em
  ``New Results from HERA''}, Schlo{\ss} Ringberg, Germany, 1999. {\tt
  hep-ph/9909310}.

\bibitem{cit:beam2}
E.~Norrbin and T.~Sj{\"o}strand, in \emph{Monte Carlo Generators for HERA
  physics}, edited by T.~A. Doyle, G.~Grindhammer, G.~Ingelman and H.~Jung, p.
  506, Proceedings of the 1998 Workshop {\em``Monte Carlo Generators for HERA
  Physics''}, Hamburg, Germany, DESY, 1999.

\bibitem{cit:zphe}
J.~{Breitweg et al., ZEUS Collab.}, Eur. Phys. J. \textbf{C6} (1999) 67.

\bibitem{cit:haykra}
M.~E. Hayes and M.~Kramer, J. Phys. \textbf{G25} (1999) 1477.

\bibitem{cit:eisen}
Y.~{Eisenberg, ZEUS Coll.}, in \emph{DIS '99}, edited by J.~Bl{\"u}mlein and
  T.~Riemann, p. 406, Proceedings of the {\em ``7th International Workshop on
  Deep Inelastic Scattering and QCD, DIS '99''}, Zeuthen, Germany, 1999,. Nucl.
  Phys. {\bf B} (Proc. Suppl.) {\bf 79} (1999) October 1999.

\bibitem{cit:blhc}
P.~{Nason et al., LHC Bottom Production Working Group}, \emph{Bottom
  Production}, Report~{\tt hep-ph/0003142}, CERN, 2000. To appear in the
  Proceedings of the {\em ``1999 CERN Workshop on Standard Model Physics (and
  more) at the LHC (First Plenary Meeting)''}.

\bibitem{cit:baar2}
M.~M. Baarmand, \emph{$b$-Quark Production at Tevatron}, Talk given at the {\em
  ``1999 CERN Workshop on Standard Model Physics (and more) at the LHC (First
  Plenary Meeting) -- LHC Bottom Production Working Group''}.

\bibitem{cit:h1beau}
C.~{Adloff et al., H1 Collab.}, Phys. Lett. \textbf{B467} (1999) 156.

\bibitem{cit:aroma}
G.~Ingelman, J.~Rathsman and G.~A. Schuler, Comp. Phys. Comm. \textbf{101}
  (1997) 135.

\bibitem{cit:bh1d}
P.~{Newman, H1 Coll.}, in \emph{DIS '99}, edited by J.~Bl{\"u}mlein and
  T.~Riemann, p. 413, Proceedings of the {\em ``7th International Workshop on
  Deep Inelastic Scattering and QCD, DIS '99''}, Zeuthen, Germany, 1999,. Nucl.
  Phys. {\bf B} (Proc. Suppl.) {\bf 79} (1999) October 1999.

\bibitem{cit:bzd}
M.~{Wing, ZEUS Coll.}, in \emph{DIS '99}, edited by J.~Bl{\"u}mlein and
  T.~Riemann, p. 416, Proceedings of the {\em ``7th International Workshop on
  Deep Inelastic Scattering and QCD, DIS '99''}, Zeuthen, Germany, 1999,. Nucl.
  Phys. {\bf B} (Proc. Suppl.) {\bf 79} (1999) October 1999.

\bibitem{cit:baar1}
M.~M. Baarmand, \emph{Recent Tevatron Results on $b$ Quark Production}, Report
  FER\-MILAB-Conf-99/145-E, Fermi National Accelerator Laboratory, 1999. To
  appear in the Proceedings of the {\em ``XXXIVth Rencontres de Moriond, QCD
  and High Energy Hadronic Interactions''}, Les Arcs, France, 1999.

\bibitem{cit:newco}
A.~P. Contogouris, Z.~Merebashvili and G.~Grispos, \emph{Polarized
  Photoproduction of Heavy Quarks in Next-to-Leading Order}, Report
  MCGILL-00-02 and UA/NPPS-01-00, McGill University and University of Athens,
  2000. {\tt hep-ph/0003204}.

\bibitem{cit:duke}
A.~Devoto and D.~W. Duke, Riv. Nuovo Cim. \textbf{7} (1984) 1.

\bibitem{cit:erdelyi}
A.~Erd\'{e}lyi, W.~Magnus, F.~Oberhettinger and F.~G. Tricomi, \emph{Higher
  Transcendental Functions}, McGraw-Hill, 1953.

\bibitem{cit:vcolor}
J.~Vermaseren, in \emph{Perturbative and Nonperturbative Aspects of Quantum
  Field Theory}, edited by H.~Latal and W.~Schweiger, p. 255, Proceedings of
  the {\em ``35.\ Internationale Universit{\"a}tswochen f{\"u}r Kern- und
  Teilchenphysik''}, Schladming, Austria, Springer, 1997.

\bibitem{cit:gellmann}
M.~Gell-Mann, Phys. Rev. \textbf{125} (1962) 1067.

\bibitem{cit:llll}
C.~H. {Llewellyn Smith}, in \emph{Quantum Flavordynamics, Quantum
  Chromodynamics, and Unified Theories}, edited by K.~T. Mahanthappa and
  J.~Randa, p.~59, Proceedings of the {\em 1979 Nato Advanced Study Institute
  at the University of Colorado}, Boulder, USA, Plenum Press, 1979.

\end{thebibliography}
\clearemptydoublepage
%
{\thispagestyle{empty}
\chapter*{Acknowledgements}

It is a pleasure to thank Prof.\ Dr.\ E.\ Reya for suggesting this
very interesting research topic to me. His generous support and his helpful 
advice have kept me on track through the years. I have benefitted
continuously from his ability to spot what is important.

I am also particularly grateful to Dr.\ M.\ Stratmann for the fruitful
collaboration on this project. He has generously shared
his insight with me and his knowledge has guided this investigation.
It is amazing how well we have worked together, in spite of communicating 
only by email most of the time.

My colleagues next door, Dr.\ S.\ Kretzer and I.\ Schienbein, have
sacrificed many hours of their valuable time for physics discussions with me. 
The free exchange of ideas with them has been very helpful and I
enjoyed it immensely.  
I would also like to thank the last three persons for critically
reading parts of the manuscript and J.\ Noritzsch for some \LaTeX\ 
magic that improved its looks. 

Finally, I would like express my gratitude 
to  Prof.\ Dr.\ M.\ Gl{\"u}ck and all the
other members of TIV for creating such an inspiring
work environment. The sheer number of talented, successful and 
open (former) TIV members I have met speaks for itself.
}
\clearpage
\end{document}